\documentstyle[12pt]{article}

\input epsf

\ifx\epsffile\undefined
\newlength{\epsfysize}
\def\epsffile#1#2#3#4]#5{}
\fi

\begin{document}

\newcommand{\be}{\begin{equation}}
\newcommand{\ee}{\end{equation}}
\newcommand{\bear}{\begin{eqnarray}}
\newcommand{\eear}{\end{eqnarray}}
\newcommand{\gsim}{\lower.7ex\hbox{$\;\stackrel{\textstyle>}{\sim}\;$}}
\newcommand{\lsim}{\lower.7ex\hbox{$\;\stackrel{\textstyle<}{\sim}\;$}}
\newcommand{\dr}{\mbox{\footnotesize{$\overline{\rm DR}$\ }}}
\newcommand{\GeV}{{\rm GeV}}
\newcommand{\TeV}{{\rm TeV}}
\newcommand{\met}{\not\!\!\!E_T}
\newcommand{\msusy}{M_{\rm SUSY}}

\def\cpc#1 #2 #3 #4 {Comp.~Phys.~Comm.  {\bf  #1}, #2 (#3)#4 }
\def\npb#1 #2 #3 #4 {Nucl.~Phys. {\bf B#1}, #2 (#3)#4 }
\def\plb#1 #2 #3 #4 {Phys.~Lett. {\bf B#1}, #2 (#3)#4 }
\def\prd#1 #2 #3 #4 {Phys.~Rev.  {\bf D#1}, #2 (#3)#4 }
\def\prl#1 #2 #3 #4 {Phys.~Rev.~Lett. {\bf #1}, #2 (#3)#4 }
\def\pr#1  #2 #3 #4 {Phys.~Rept. {\bf #1}, #2 (#3)#4 }
\def\mpl#1 #2 #3 #4 {Mod.~Phys.~Lett. {\bf A#1}, #2 (#3)#4 }
\def\zpc#1 #2 #3 #4 {Z.~Phys. {\bf C#1}, #2 (#3)#4 }
%

\pagestyle{empty}
\begin{titlepage}
\def\thepage {}        

\title{\bf Tau Jet Signals for Supersymmetry\\
            at the Tevatron\\ [1cm]}

\author{ {\bf Joseph~D.~Lykken and Konstantin~T.~Matchev\thanks{Based on
talks given at the 
Summary Meeting of the SUSY/Higgs Run II Workshop,
Fermilab,  November 19-21, 1998; the
{\em ``Higgs and Supersymmetry: Search and Discovery''} conference,
Gainesville, FL, March 7-11, 1999; and the
Pheno'99 Symposium: {\em ``Phenomenology for the Third Millenium''},
Madison, WI,  April 12-14, 1999.}} \\ 
\\
{\small {\it Theoretical Physics Department}}\\  
{\small {\it Fermi National Accelerator Laboratory}}\\
{\small {\it Batavia, Illinois, 60510, USA \thanks{e-mail
  addresses: lykken@fnal.gov, matchev@fnal.gov} }}\\ }

\date{ }

\maketitle

   \vspace*{-9.9cm}
\noindent

\rightline{FERMILAB-Conf-99-270-T}  
\rightline{October 1, 1999}    

\vspace*{9.4cm}

\baselineskip=18pt

\begin{abstract}

{\normalsize  We present a more detailed account of our study
(hep-ph/9903238) for the supersymmetry reach of the Tevatron in
channels with isolated leptons and identified tau jets.
We review the theoretical motivations for expecting such
signatures, and describe the relevant parameter space in
the minimal supergravity and the minimal gauge-mediated models.
With explicit Monte Carlo simulations we then show that
for certain parameter ranges, channels with two leptons and
one tau jet offer a better reach in Run II than the clean
trilepton signal. We emphasize than improving on tau ID
is an important prerequisite for successful searches in
multiple tau jet channels. Finally, we discuss some triggering issues.}

\end{abstract}

\vfill
\end{titlepage}

\baselineskip=18pt
\pagestyle{plain}
\setcounter{page}{1}

\section{Introduction and Motivation}\label{sec:introduction}

No matter how one looks at it, the third generation in the Standard Model (SM)
is special. (The 3rd generation fermions may provide a clue to the origin of mass,
fourth generation, etc.) This is even more so in the Minimal Supersymmetric
Standard Model (MSSM), where the third generation superpartners
are singled out in several ways. First, their larger Yukawa couplings
tend to drive the corresponding soft scalar masses smaller through
the RGE evolution. Second, they play an important role in triggering
radiative electroweak symmetry breaking, and as a result, fine-tuning
arguments suggest that they are probably lighter than the other two
generations\footnote{Notice, however, that unlike the first two generations,
their masses are not so well restricted by the stringent constraints
coming from flavor-changing processes.} \cite{DG}.
Finally, squark and slepton mixing 
for the third generation is typically rather large and further decreases
the mass of the lightest mass eigenstates.
For all of these reasons, it is possible that the third generation squarks
and sleptons could be relatively light and therefore more easily accessible
at the current and future colliders. Then, a logical thing to do will be
to study particular signatures involving their decay products -
top and bottom quarks, and tau leptons or neutrinos. Of these four,
the tau leptons appear as the most promising possibility at the Tevatron.
Tau neutrinos are invisible, and they often come paired with
tau leptons anyway. Signatures with b-jets are also promising, but
they tend to have large QCD backgrounds. And finally, top quarks
are heavy, which limits the Tevatron reach for those channels.

Searches for supersymmetry (SUSY) in Run I of the Tevatron
have been done exclusively in channels involving some
combination of leptons, jets, photons and missing transverse
energy ($\met$) \cite{Tevatron searches}. 
At the same time, several Run I analyses have identified hadronic tau
jets in the most abundant Standard Model processes, e.g. in $W$-production
\cite{W to tau} and top decays \cite{top to tau}.
Hadronically decaying taus have also been used to place limits on
a charged Higgs \cite{H+ to tau} and leptoquarks \cite{Leptoquark to tau}.
Since tau identification is expected to improve further in Run II,
this raises the question whether SUSY searches in channels involving
tau jets are feasible.

SUSY signatures with tau leptons are very well motivated, since
they arise in a variety of models of low-energy supersymmetry, e.g.
gravity mediated (SUGRA) \cite{BCDPTinPRL,BCDPTinPRD,JW}
or the minimal gauge mediated (MGM) models \cite{JW,DN,BT}.
Here we present results from a study \cite{LM} of all possible
{\em experimental} signatures with three identified objects
(leptons or tau jets) plus $\met$, and compare their reach
to the clean trilepton channel \cite{Run I 3L,3L,Barger,BCDPTinPRL}.
In evaluating the physics potential of the future Tevatron
runs in these new tau channels, it is important to
be aware not only of the physical backgrounds, but also of the 
experimental realities. Jets faking taus will comprise a significant
fraction of the background, and it is crucial to have a reliable
estimate of that rate, which we attempt to estimate from
a detailed Monte Carlo analysis.
We used PYTHIA \cite{PYTHIA} and TAUOLA \cite{tauola}
for event generation, and the SHW package \cite{SHW},
which provides a realistic Run II detector simulation.

In the next Section we delineate the relevant parameter space regions
of the minimal SUGRA and MGM models, where one may expect enhanced
tau signals. We then discuss in rather general terms the pros
and cons of the tau jet channels.
Later in Section~\ref{sec:analysis} we describe in
detail our analysis and present our cut selection. In
Section~\ref{sec:backgrounds} we discuss the major SM backgrounds,
and in Section~\ref{sec:triggering} we perform a study 
on triggering in those new channels.
Finally in Section~\ref{sec:reach} we show the expected
Run II Tevatron reach for the scenario under consideration.

\section{Tau Signals in SUGRA and MGM Models}\label{sec:models}

Most people would probably agree that our best bet to discover
supersymmetry at the Tevatron is the clean $3\ell\met$ channel.
It arises in the decays of gaugino-like 
chargino-neutralino pairs $\tilde\chi^\pm_1 \tilde \chi^0_2 $. 
The reach is somewhat limited by the rather small leptonic
branching fractions of the chargino and neutralino. In the limit
of either heavy or equal in mass squarks and sleptons,
the leptonic branching ratios are $W$-like
and $Z$-like, respectively. However, both
gravity mediated and gauge mediated models of SUSY breaking
allow the sleptons to be much lighter than the squarks,
thus enhancing the leptonic branching fractions of the gauginos.
What is more, in certain regions of parameter space the lightest
tau slepton can be much lighter than the other sleptons, and
then the gaugino decays will proceed predominantly to final
states with tau leptons only.

\subsection{Light sleptons}

There are various generic reasons as to why one may expect
light sleptons in the spectrum. For example, the slepton masses
at the high-energy (GUT or messenger) scale may be rather small
to begin with. This is typical for gauge mediated models,
since the sleptons are colorless and do not receive large
soft mass contributions $\sim \alpha_s$. 
The minimal SUGRA models, on the other hand, predict light sleptons
in the region of parameter space where $M_0\ll M_{1/2}$.
Various effects (non-flat Kahler metric, RGE running above the
GUT scale, D-terms from extra $U(1)$ gauge factors)
may induce nonuniversalities in the scalar masses at the GUT scale,
in which case the slepton-squark mass hierarchy can be affected.
In the absence of a specific model, we do not know which way the
splittings will go, but as long as the soft scalar masses are small,
the RGE running down to the weak scale will naturally induce a
splitting between the squarks and sleptons, making the sleptons lighter.
Now, given that the sleptons are the lightest scalars in the spectrum,
it is quite plausible that by far the lightest among them are
the third generation sleptons. As we mentioned in
Section~\ref{sec:introduction}, RGE running and mixing in the charged
slepton sector may push the stau masses down.

As a result of some or all of these effects, it may 
very well be that among all scalars, only the lightest sleptons
from each generation (or maybe just the lightest stau $\tilde\tau_1$)
are lighter than $\tilde\chi^\pm_1$ and $\tilde \chi^0_2$.
Indeed, in both SUGRA and minimal gauge mediated models
one readily finds regions of parameter space where either 
$$
m_{\tilde\chi^0_1}< m_{\tilde \tau_1}
\sim m_{\tilde \mu_R}< m_{\tilde\chi^+_1}
$$
(typically at small $\tan\beta$) or
$$
m_{\tilde\chi^0_1}< m_{\tilde \tau_1}
< m_{\tilde\chi^+_1}< m_{\tilde \mu_R}
$$
(at large $\tan\beta$). Depending on the particular model, and the
values of the parameters, the gaugino pair decay chain may then end
up overwhelmingly in {\em any one} of the four final states:
$\ell\ell\ell$, $\ell\ell\tau$, $\ell\tau\tau$ or $\tau\tau\tau$. 
(From now on, we shall use the following terminology: a ``lepton''
($\ell$) is either a muon or an electron; a tau ($\tau$) is a tau-lepton,
which can later decay either leptonically, or to a hadronic tau jet,
which we denote by $\tau_h$).

\begin{figure}[t!]
\epsfysize=3.0in
\epsffile[-50 230 425 580]{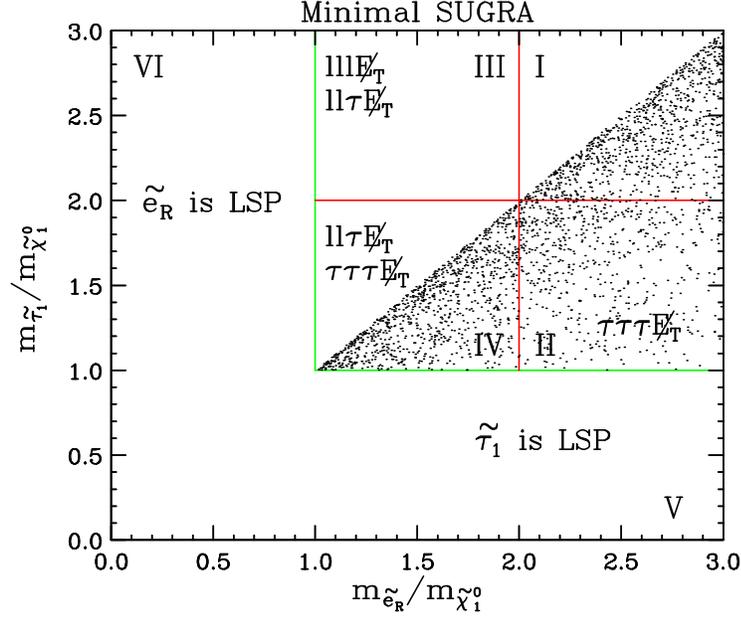}
\begin{center}
\parbox{5.5in}{
\caption[]
{\small Scatter plot of minimal SUGRA model points versus the ratios 
$m_{\tilde e_R}/m_{\tilde\chi_1^0}$ and 
$m_{\tilde\tau_1}/m_{\tilde\chi_1^0}$.
\label{SUGRA}}}
\end{center}
\end{figure}

In Fig.~\ref{SUGRA} we show a scatter plot\footnote{Notice that
the plots in this report are best viewed on a color screen,
or when printed on a color printer.} of SUGRA model points
plotted versus the ratios $m_{\tilde e_R}/m_{\tilde\chi_1^0}$ and 
$m_{\tilde\tau_1}/m_{\tilde\chi_1^0}$ (see \cite{PBMZ} for details
on how the sampling was done). 
We concentrate 
on the region $M_0\ll M_{1/2}$ and show only points with
$m_{\tilde\tau_1}<3m_{\tilde\chi_1^0}$ and 
$m_{\tilde e_R  }<3m_{\tilde\chi_1^0}$.
(Note that in SUGRA the lightest
selectron or smuon is purely right-handed, while the lightest stau
typically has a sizable left-handed component.)
There are several distinct regions in relation to the branching
ratios of the chargino-neutralino pair (recall that in SUGRA 
$m_{\tilde\chi^+_1}\sim m_{\tilde\chi^0_2}
\sim 2m_{\tilde\chi^0_1}$):
\begin{itemize}
\item Region I:
$m_{\tilde e_R}   > m_{\tilde\chi^+_1}$ and
$m_{\tilde \tau_1}> m_{\tilde\chi^+_1}$.
In this case, all two-body decays are closed, and the leptonic
branching ratios of the gauginos are $W$-like ($Z$-like).
\item Region II:
$m_{\tilde e_R} > m_{\tilde\chi^+_1}$, but
$m_{\tilde\chi^0_1}< m_{\tilde \tau_1}< m_{\tilde\chi^+_1}$,
so that 
$BR(\tilde\chi^+_1\tilde\chi^0_2\rightarrow \tau\tau\tau)\simeq 100\%$.
Note that if the stau mass is too close to either $m_{\tilde\chi^0_1}$
or $m_{\tilde\chi^+_1}$, at least one of the resulting taus 
will be quite soft. One would therefore expect the largest efficiency
if $m_{\tilde\tau_1}\simeq (m_{\tilde\chi^+_1}+m_{\tilde\chi^0_1})/2$.
\item
Region III:
$m_{\tilde \tau_1} > m_{\tilde\chi^+_1}$ and
$m_{\tilde\chi^0_1}< m_{\tilde e_R}< m_{\tilde\chi^+_1}$.
Then the gauginos can only decay to selectrons or smuons
via two-body decays. Note that $\tilde\chi_2^0$ is mostly $\tilde W_3$,
while $\tilde \chi^+_1$ is mostly $\tilde W^+$, and those do not couple
to right-handed squarks or sleptons. Therefore the decay
$\tilde\chi^0_2\rightarrow \tilde \ell^\pm \ell^\mp$ proceeds through
the relatively small $\tilde B$ component of the $\tilde\chi^0_2$,
while the decay 
$\tilde\chi^+_1\rightarrow \tilde \ell^+ \nu_\ell$
is severely suppressed by the small muon or electron Yukawa couplings,
and the three-body decays 
$\tilde\chi^+_1\rightarrow \tilde\chi_1^0 \ell^+ \nu_\ell$,
$\tilde\chi^+_1\rightarrow \tilde\chi_1^0 \tau^+ \nu_\tau$
become dominant. Since those can also be mediated
by an off-shell $W$, we expect both of them to be present.
Notice how the assumption of generational independence
of the scalar masses at the GUT scale assures that
$m_{\tilde \tau_1} < m_{\tilde \ell_R}$, so that there are no
SUGRA model points in region III, but this can be avoided if one
alows for different stau and first two generation slepton
masses at the GUT scale \cite{AAD}\footnote{Such a situation,
however, is not well motivated from the point of view of
SUSY GUTs. One can imagine that strict universality holds
at the Planck scale, and then RGE running down to the
GUT scale introduces intergenerational mass splittings.
But then, due to the large tau Yukawa coupling,
we would expect the tau slepton masses to be
the lightest slepton masses at the GUT scale.}.
\item
Region IV:
$m_{\tilde\chi^0_1}< m_{\tilde e_R}< m_{\tilde\chi^+_1}$
and
$m_{\tilde\chi^0_1}< m_{\tilde \tau_1}< m_{\tilde\chi^+_1}$,
so that the signatures from both regions II and III can be present.
Now, the trilepton signal is somewhat suppressed,
since the chargino decays mostly to taus.
\item 
Region V:
$m_{\tilde \tau_1}< m_{\tilde\chi^0_1}$.
Here one finds a charged LSP (stau), which is
stable, if R-parity is conserved, and therefore excluded
cosmologically.
\item
Region VI:
$m_{\tilde e_R}< m_{\tilde\chi^0_1}$. 
This region is excluded for the same reason as Region V,
since now the smuon is the LSP.
\end{itemize}
To summarize, in SUGRA models, on most general grounds we expect
chargino-neutralino pair production to give rise to 
$\tau\tau\tau$, $\tau\ell\ell$ or $\ell\ell\ell$ final states, where
the first two can be dominant in certain regions of parameter space.

\begin{figure}[t!]
\epsfysize=3.0in
\epsffile[-50 230 425 580]{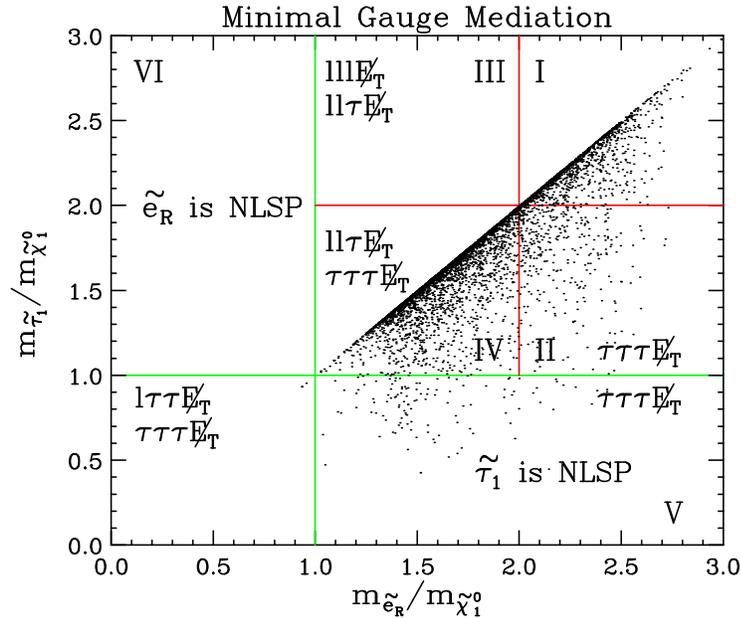}
\begin{center}
\parbox{5.5in}{
\caption[]
{\small Scatter plot of MGM model points versus the ratios 
$m_{\tilde e_R}/m_{\tilde\chi_1^0}$ and 
$m_{\tilde\tau_1}/m_{\tilde\chi_1^0}$.
\label{GM}}}
\end{center}
\end{figure}

We next consider the minimal gauge mediated models
(we follow the conventions of Ref.~\cite{BMPZinPRD})
and show the corresponding scatter plot in Fig.~\ref{GM}. 
Our discussion of regions I-IV above applies here as well.
The novel feature is that now the goldstino $\tilde G$ is the LSP,
and therefore regions V and VI are in principle allowed.
We do indeed find points in those regions, but only if
$m_{\tilde e_R}>m_{\tilde\tau_1}$. This is again a
consequence of the generation independence of the scalar masses
at the messenger scale, which is a robust prediction of the
minimal gauge-mediated models. In order to avoid this argument,
one would have to allow for messenger-matter mixing and arrange
for different couplings of the messengers to the three families.

A very interesting situation may arise in the intersection of regions
V and VI. If the mass splitting between 
$\tilde e_R$, $\tilde\mu_R$ and ${\tilde\tau_1}$ is very small
(i.e. at rather small values of $\tan\beta$), they may all be co-NLSP's.
Just as before, $\tilde\chi^+_1$ will preferentially decay to taus:
$\tilde\chi^+_1\rightarrow \tau \nu_\tau \tilde G$.
The neutralino decays, however, are of two sorts:
$\tilde\chi^0_i \rightarrow \tilde\tau^\pm_1 \tau^\mp
\rightarrow \tau^\pm\tau^\mp \tilde G$
and 
$\tilde\chi^0_i \rightarrow \tilde \ell_R^\pm \ell^\mp
\rightarrow \tilde \tau^\pm_1 \ell^\mp X
\rightarrow \tau^\pm \ell^\mp \tilde G X$, where
$i=1,2$ and $X$ stands for the very soft products of the selectron
(or smuon) decay to a stau. The typical signature in this
case would be $\tau\tau\ell$.

\subsection{Tau Jets}

The above discussion of the two most popular supersymmetric models
reveals that, depending on the model parameters,
the gaugino decay chains may overwhelmingly end up in any one of
the four final states $\tau\tau\tau$, $\tau\tau\ell$, $\tau\ell\ell$
and $\ell\ell\ell$. In order to decide as to which experimental signatures
are most promising, we have to first factor in the tau branching ratios
to leptons\footnote{Recall that here we call only the electrons and muons
``leptons'', following experimentalists' lingo.} and jets.
About two-thirds of the subsequent tau decays are hadronic,
so it appears advantageous to consider signatures with tau jets
in the final state as alternatives to the clean trilepton signal.
The branching ratios for three leptons or undecayed taus into a final state
containing leptons and tau jets is shown in Table~\ref{tau_branching}.
\begin{table}[t!]
\centering
\renewcommand{\arraystretch}{1.5}
\begin{tabular}{||c||c|c|c|c||}
\hline\hline
Experimental
  & \multicolumn{4}{c||}{Trilepton SUSY signal} \\ 
\cline{2-5}
signature
  & $\tau\tau\tau$ 
    & $\tau\tau\ell$
      & $\tau\ell\ell$
        & $\ell\ell\ell$  \\
\hline \hline
$\tau_h\tau_h\tau_h$
  & 0.268
    & ---
      & ---
        & ---  \\
\hline
$\ell\tau_h\tau_h$
  & 0.443
    & 0.416
      & ---
        & ---  \\
\hline
$\ell\ell\tau_h$
  & 0.244
    & 0.458
      & 0.645
        & ---  \\
\hline
$\ell\ell\ell$ 
  & 0.045
    & 0.126
      & 0.355
        & 1.00  \\
\hline\hline
\end{tabular}
\parbox{5.5in}{
\caption{ Branching ratios of the four possible
SUSY signals into the corresponding
experimental signatures involving final state leptons 
$\ell$ (electrons or muons) as well as identified
tau jets ($\tau_h$). \label{tau_branching}}}
\end{table}
We see that the presence of taus in the underlying SUSY signal
always leads to an enhancement of the signatures with tau jets
in comparison to the clean trileptons. This disparity is most striking
for the case of $\tau\tau\tau$ decays, where
$BR(\tau\tau\tau\rightarrow \ell\ell\tau_h)/
 BR(\tau\tau\tau\rightarrow \ell\ell\ell)\sim 5.5$.

\begin{figure}[t!]
\epsfysize=3.0in
\epsffile[-70 225 280 555]{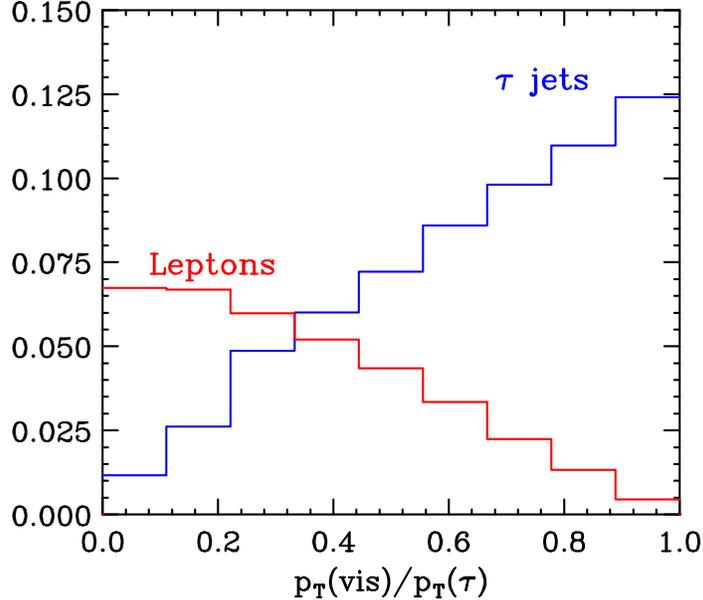}
\begin{center}
\parbox{5.5in}{
\caption[]
{\small Distribution of the $p_T$ fraction that the visible tau
decay products (charged leptons or tau jets) inherit from the tau parent.
\label{pT_frac}}}
\end{center}
\end{figure}

An additional advantage of the tau jet channels over the clean trileptons
is that the leptons from tau decays are much softer than the tau jets
and as a result will have a relatively low reconstruction efficiency.
We illustrate this point in Fig.~\ref{pT_frac}, where we show
the distribution of the $p_T$ fraction carried away by the visible
decay products (charged lepton or tau jet) in tau decays
(for theoretical discussions, see \cite{tau decays}).
We can see that the leptons from tau decays are very soft,
and it has been suggested \cite{Barger} to use softer lepton
$p_T$ cuts in order to increase signal acceptance.

\begin{figure}[t!]
\epsfysize=3.0in
\epsffile[-70 225 280 555]{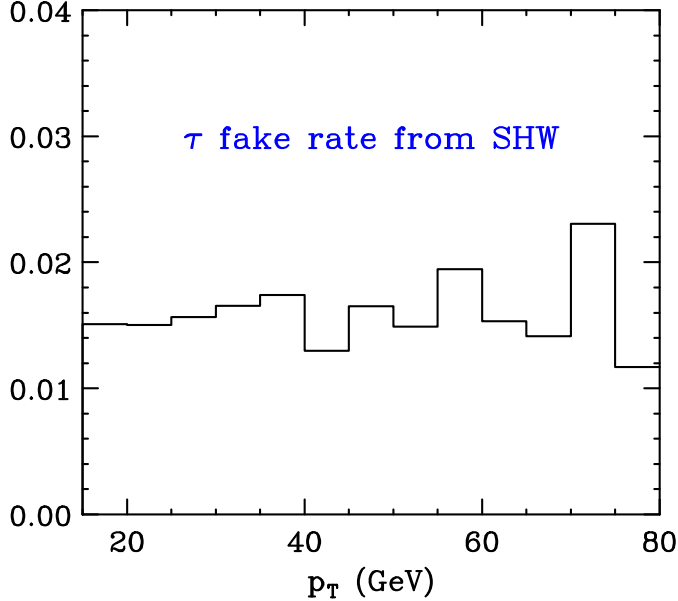}
\begin{center}
\parbox{5.5in}{
\caption[]
{\small The tau fake rate defined as the number of QCD jets
misidentified as taus over the total number of reconstructed
QCD jets, in $W$ events.
\label{tau fake rate}}}
\end{center}
\end{figure}

However, there are also some factors, which work against the
tau jet channels. First and foremost, the background 
in those channels is larger than for the clean trileptons.
The physical background (from {\em real} tau jets in the event)
is actually smaller, but a significant part of the background is due to
events containing narrow isolated QCD jets with the correct track
multiplicity, which can be misidentified as taus. 
In Fig.~\ref{tau fake rate} we show the tau fake rate that we obtained
from SHW in $W$ events.
We define the fake rate as the number of QCD jets misidentified
as taus over the total number of reconstructed QCD jets.
The fake rate that we find with SHW is somewhat higher than
in real data and/or with full CDF detector simulation
\cite{Hohlmann,Leslie}. This is to be expected in a much cleaner simulated
environment, where, unlike real data, there is less junk flying around,
and the jets tend to pass the isolation cuts more easily.

The jetty signatures are also hurt by
the lower detector efficiency for tau jets than for leptons.
The main goal of our study, therefore, was to see what would be the
net effect of all these factors, on a channel by channel basis.

\subsection{A Challenging Scenario}

For our analysis we choose to examine one of the most challenging
scenarios for SUSY discovery at the Tevatron.
We assume the typical large $\tan\beta$ mass hierarchy
$m_{\tilde\chi^0_1}< m_{\tilde \tau_1}<
m_{\tilde\chi^+_1}< m_{\tilde \mu_R}$. One then finds that
$BR(\tilde\chi^+_1\tilde\chi^0_2\rightarrow \tau\tau\tau+X)\simeq 100\%$
below $\tilde\chi^\pm_1\rightarrow W^\pm \tilde\chi^0_1$
and   $\tilde\chi^0_2\rightarrow Z\tilde\chi^0_1$ thresholds.
In order to shy away from specific model dependence, we shall
conservatively ignore all SUSY production channels other than
$\tilde\chi^\pm_1\tilde \chi^0_2 $ pair production.
The $p_T$ spectrum of the taus resulting from the chargino and
neutralino decays depends on the mass differences
$m_{\tilde\chi^+_1}-m_{\tilde\tau_1}$ and
$m_{\tilde\tau_1}-m_{\tilde\chi^0_1}$.
The larger they are, the harder the spectrum, and the better the
detector efficiency. However, as the mass difference gets large,
the $\tilde\chi^+_1$ and $\tilde\chi^0_2$ masses themselves
become large too, so the production cross-section is severely suppressed.
Therefore, at the Tevatron we can only explore regions with favorable
mass ratios and at the same time small enough gaugino masses.
This suggests a choice of SUSY mass ratios: for definiteness we fix
$2m_{\tilde\chi^0_1}\sim (4/3)\ m_{\tilde \tau_1}
  \sim m_{\tilde\chi^+_1} (< m_{\tilde \mu_R})$
throughout the analysis, and vary the chargino mass.
The rest of the superpartners have fixed large masses
corresponding to the mSUGRA point $M_0=180$ GeV, 
$M_{1/2}=180$ GeV, $A_0=0$ GeV, $\tan\beta=44$ and $\mu>0$,
but we are not constrained to mSUGRA models only.
Our analysis will apply equally to gauge-mediated models with
a long-lived neutralino NLSP, as long as the relevant gaugino and slepton
mass relations are similar. Note that our choice of heavy
first two generation sleptons is very conservative.
A more judicious choice of their masses, namely
$m_{\tilde \mu_R}<m_{\tilde\chi^+_1}$, would lead to a
larger fraction of trilepton events, and as a result, a higher reach.
Furthermore, the gauginos would then decay via two-body modes
to first generation sleptons, and the resulting lepton spectrum
would be much harder, leading to a higher lepton efficiency.
Notice also that the $\tilde\chi^\pm_1\tilde\chi^0_2$
production cross-section is sensitive to the squark masses,
but since this is the only production process we are considering,
our results can be trivially rescaled to account for a different
choice of squark masses, or to include other production processes
as well.

\section{Analysis}\label{sec:analysis}

We used PYTHIA v6.115 and TAUOLA v2.5 for event generation.
We used the SHW v2.2 detector simulation package \cite{SHW}, which
simulates an average of the CDF and D0 Run II detector performance.
In SHW tau objects are defined as jets with $|\eta|<1.5$, net
charge $\pm 1$, one or three tracks in a $10^\circ$
cone with no additional tracks in a $30^\circ$
cone, $E_T>5$ GeV, $p_T>5$ GeV, plus an electron rejection cut.
SHW electrons are required to have $|\eta|<1.5$, $E_T>5$ GeV,
hadronic to electromagnetic energy
deposit ratio $R_{h/e}<0.125$, and satisfy standard isolation cuts.
Muon objects are required to have $|\eta|<1.5$, $E_T>3$ GeV
and are reconstructed using Run I efficiencies. We use standard isolation
cuts for muons as well. Jets are required to have $|\eta|<4$, 
$E_T>15$ GeV. In addition we have added jet energy correction
for muons and the rather loose id requirement $R_{h/e}>0.1$.
We have also modified the TAUOLA program in order to correctly
account for the chirality of tau leptons coming from SUSY decays.

The reconstruction algorithms in SHW already include some
basic cuts, so we can define a reconstruction efficiency
$\epsilon_{rec}$ for the various types of objects: electrons,
muons, tau jets etc. We find that as we vary the chargino mass
from 100 to 140 GeV the lepton and tau jet reconstruction
efficiencies for the signal
range from 42 to 49 \%, and from 29 to 36\%, correspondingly.
The lepton efficiency may seem surprisingly low, but this is because
a lot of the leptons are very soft and fail the $E_T$ cut.
The tau efficiency is in good agreement with the results from 
Ref.~\cite{Hohlmann} and \cite{Leslie}, once we account
for the different environment, as well as cuts used in those
analyses.

As we already emphasized earlier, 
the most important background issue in the new tau channels is the fake
tau rate. Several experimental analyses try to estimate it using
Run I data. Here we simulate the corresponding backgrounds to our
signal and use SHW to obtain the fake rate, thus
avoiding trigger bias \cite{Hohlmann}. 

\subsection{Cuts}

We now list our cuts for each channel.

\begin{figure}[t!]
\epsfysize=3.5in
\epsffile[0 250 480 600]{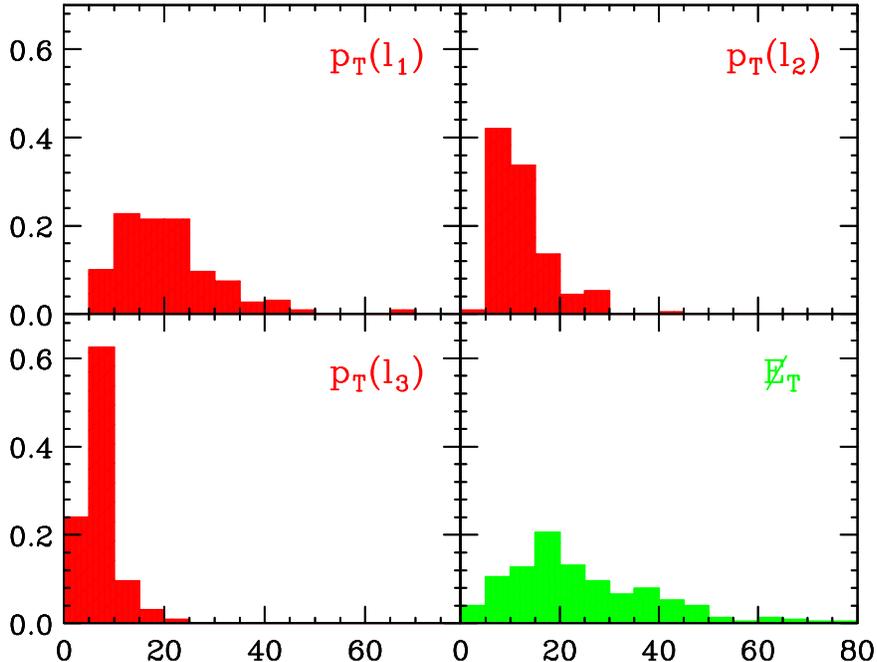}
\begin{center}
\parbox{5.5in}{
\caption[]
{\small $p_T$ distributions of the three leptons and $\met$ distribution
(all normalized to unit probability) in $\ell\ell\ell\met$ signal events,
for $m_{\tilde\chi^+_1}=123$ GeV.
\label{0T3L}}}
\end{center}
\end{figure}

As discussed earlier, we expect that the reach in the classic
$\ell\ell\ell\met$ channel will be quite suppressed, due to the
softness of the leptons (we show the $p_T$ distribution of the three
leptons, as well as the $\met$ distribution, in Fig.~\ref{0T3L}).
Therefore we apply the soft cuts advertised in Refs.~\cite{Barger}. 
We require a central lepton with 
$p_T>11$ GeV and $|\eta|<1.0$, and in addition two more leptons with
$p_T>7$ GeV and $p_T>5$ GeV.
Leptons have to be isolated: $I(\ell)<2$ GeV, where $I$ is
the total transverse energy contained in a cone of size
$\delta R=\sqrt{\Delta\varphi^2+\Delta\eta^2}=0.4$
around the lepton. 
We impose a dilepton invariant mass cut for same flavor,
opposite sign leptons: $|m_{\ell^+\ell^-}-M_Z|>10$ GeV
and $|m_{\ell^+\ell^-}|>11$.
Finally, we impose an optional veto on additional jets
and require $\met$ to be either
more than 20 GeV, or 25 GeV. This gives us a total of
four combinations of the $\met$ cut and the jet veto
(shown in Table~\ref{table cuts}),
which we apply for all tau jet signatures later as well. 
\begin{table}[t!]
\centering
\renewcommand{\arraystretch}{1.5}
\begin{tabular}{||c||c|c||}\hline\hline
Sample & $\met$ cut  & Jet veto  \\ \hline\hline
A      & 20 GeV      & no        \\ \hline
B      & 25 GeV      & no        \\ \hline
C      & 20 GeV      & yes       \\ \hline
D      & 25 GeV      & yes       \\ \hline\hline
\end{tabular}
\parbox{5.5in}{
\caption{ Definition of the signal samples A-D.
\label{table cuts}}}
\end{table}

\begin{figure}[t!]
\epsfysize=3.5in
\epsffile[0 250 480 600]{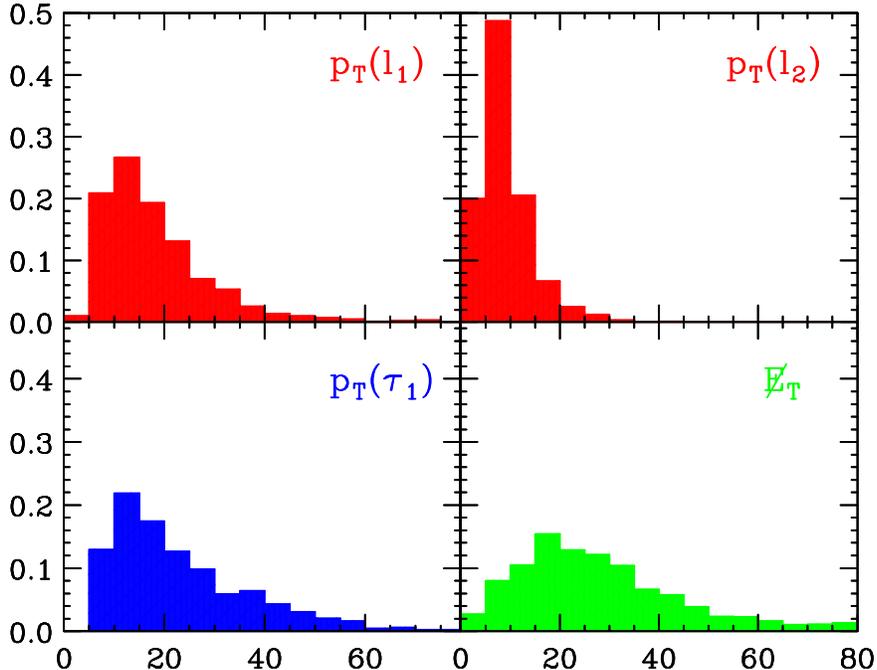}
\begin{center}
\parbox{5.5in}{
\caption[]
{\small The same as Fig.~\ref{0T3L}, but for the two leptons
and the tau jet in $\ell\ell\tau_h\met$ signal events.
\label{1T2L}}}
\end{center}
\end{figure}

For our $\ell\ell\tau_h\met$ analysis we impose cuts similar to the
stop search analysis in the $\ell^+\ell^-j\met$ channel \cite{CDFanalysis}:
two isolated ($I(\ell)<2$ GeV) leptons with $p_T>8$ GeV and $p_T>5$ GeV, and
one identified tau jet with $p_T(\tau_h)>15$ GeV (the $p_T$ and
$\met$ distributions are shown in Fig.~\ref{1T2L}).
Again, we impose the above
invariant mass cuts for any same flavor, opposite sign dilepton pair.
This channel was previously considered in Ref.~\cite{BCDPTinPRD},
but with somewhat harder cuts on the leptons.

A separate, very interesting signature ($\ell^+\ell^+\tau_h\met$)
arises if the two leptons
have the same sign, since the background is greatly suppressed.
In fact, we expect this background to be significantly smaller than the
trilepton background! Roughly one third of the signal events in
the general $\ell\ell\tau_h$ sample are expected to have like-sign leptons.

\begin{figure}[t!]
\epsfysize=3.5in
\epsffile[0 250 480 600]{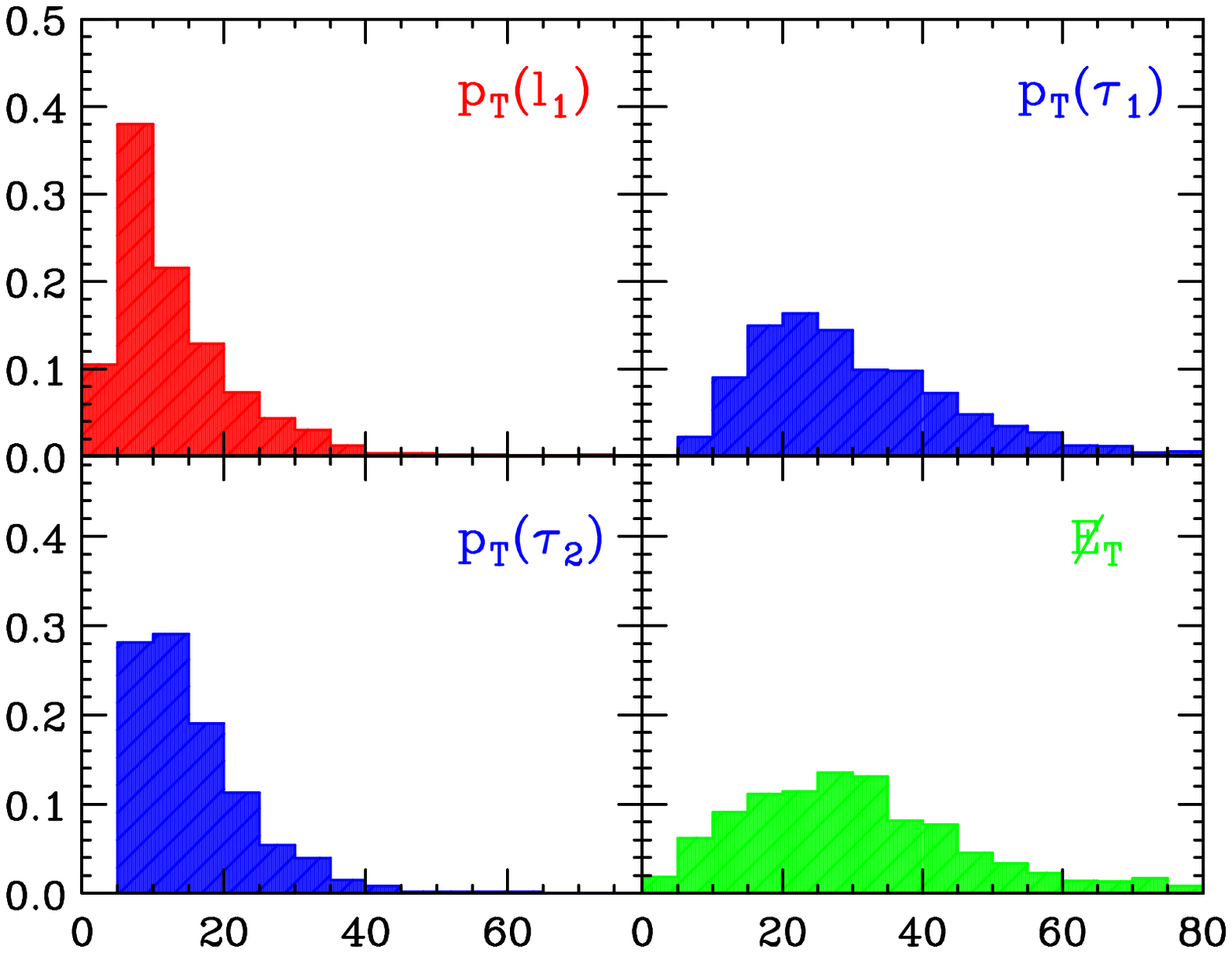}
\begin{center}
\parbox{5.5in}{
\caption[]
{\small The same as Fig.~\ref{0T3L}, but for the lepton
and the two tau jets in $\ell\tau_h\tau_h\met$ signal events.
\label{2T1L}}}
\end{center}
\end{figure}

For our $\ell\tau_h\tau_h\met$ analysis we use some basic
identification cuts: two tau jets with $p_T>15$ GeV and  $p_T>10$ GeV
and one isolated lepton with $p_T>7$ GeV. The corresponding $p_T$
and $\met$ distributions are shown in Fig.~\ref{2T1L}.

\begin{figure}[t!]
\epsfysize=3.5in
\epsffile[0 250 480 600]{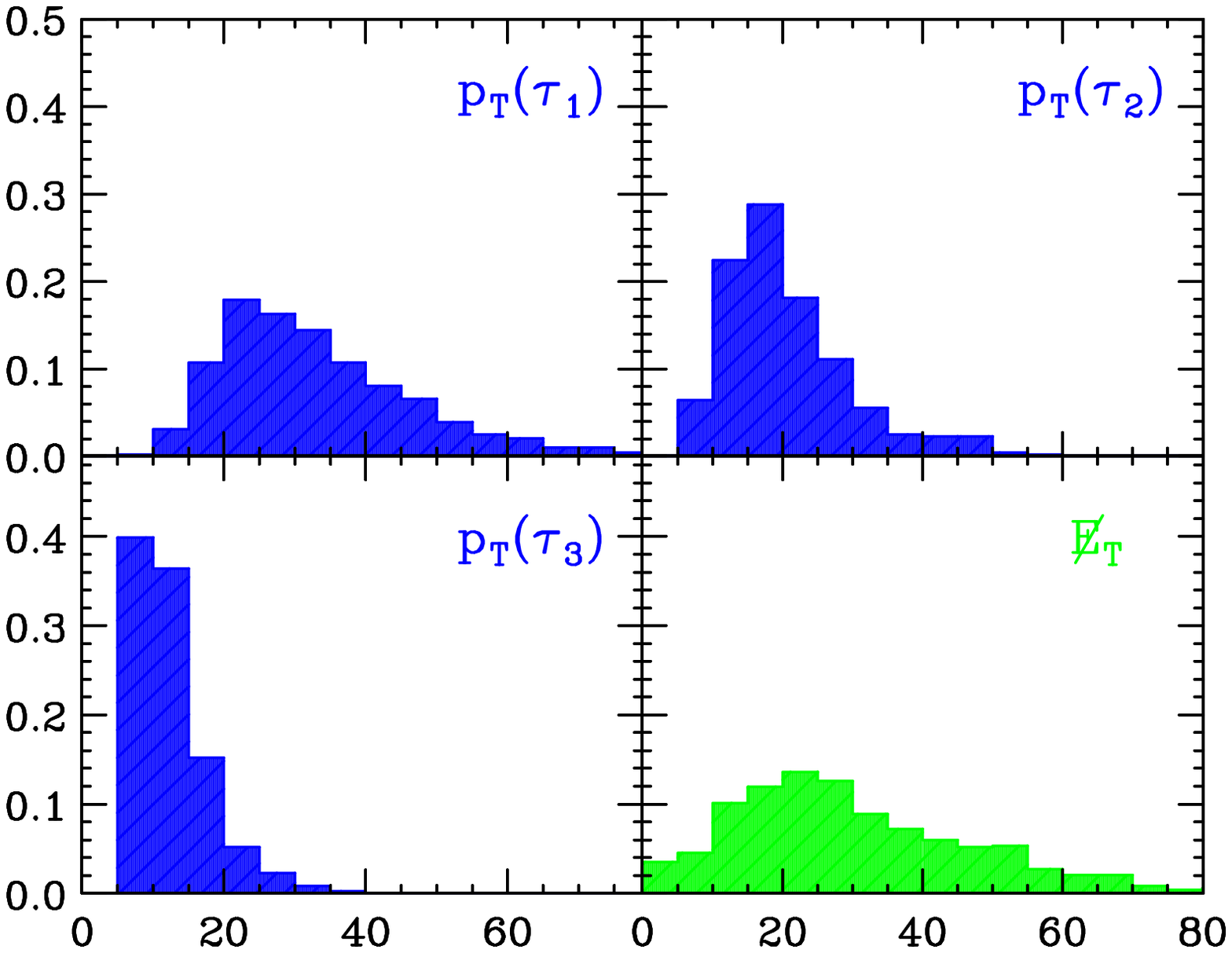}
\begin{center}
\parbox{5.5in}{
\caption[]
{\small The same as Fig.~\ref{0T3L}, but for the 
tau jets in $\tau_h\tau_h\tau_h\met$ signal events.
\label{3T0L}}}
\end{center}
\end{figure}

Finally, for the $\tau_h\tau_h\tau_h\met$ signature
we only require three tau jets with
$p_T>15,10$ and 8 GeV, respectively (Fig.~\ref{3T0L}).

\subsection{Signal}

\begin{figure}[t!]
\epsfysize=3.0in
\epsffile[-80 215 240 600]{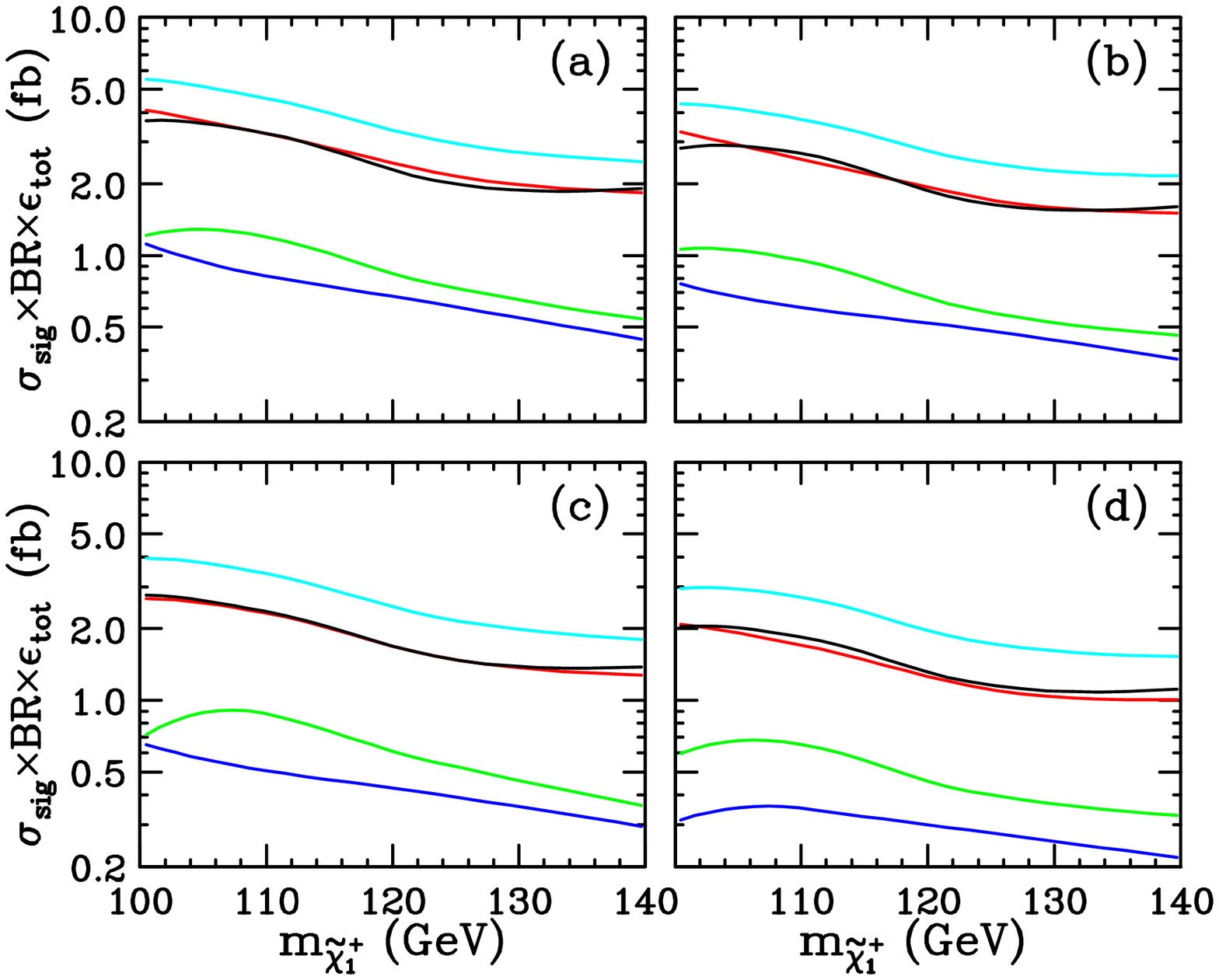}
\begin{center}
\parbox{5.5in}{
\caption[]{ Signal cross-section times branching ratio after cuts
for the five channels discussed in the text:
$\ell\ell\ell\met$ (blue),
$\ell\ell\tau_h\met$ (red),
$\ell^+\ell^+ \tau_h\met$ (green),
$\ell\tau_h\tau_h\met$ (cyan) and
$\tau_h\tau_h\tau_h\met$ (black);
and for various sets of cuts: (a) cuts A,
(b) cuts B, (c) cuts C and (d) cuts D.
\label{sigeff}}}
\end{center}
\end{figure}

One can get a good idea of the relative importance of the
different channels by looking at the corresponding signal samples
after the analysis cuts have been applied.
In Fig.~\ref{sigeff} we show the signal cross-sections
times the corresponding branching ratios times the total efficiency
$\epsilon_{tot}\equiv \epsilon_{rec}\epsilon_{cuts}$,
which accounts for both the detector acceptance $\epsilon_{rec}$
and the efficiency of the cuts $\epsilon_{cuts}$ (for each signal
point we generated $10^5$ events).
We see that the lines are roughly ordered according to the
branching ratios from Table~\ref{tau_branching}. This can be
understood as follows. The acceptance (which includes the
basic ID cuts in SHW) is higher for leptons than for $\tau$ jets.
Therefore, replacing a lepton with a tau jet in the
experimental signature costs us a factor of $\sim 1.5$ in acceptance,
due to the poorer reconstruction of tau jets, compared to leptons.
Later, however, the cuts tend to reduce the leptonic signal more than the
tau jet signal. This is mostly because the leptons are softer
than the tau jets.
Notice that we cannot improve the efficiency for leptons by further lowering
the cuts -- we are already using the most liberal cuts \cite{Barger}.
It turns out that these two effects mostly cancel each other, and the total
efficiency $\epsilon_{tot}$ is roughly the same for all channels.
Therefore the relative importance of each channel
will only depend on the tau branching ratios and the backgrounds. 
For example, in going from $\ell\ell\ell$ to $\ell\ell\tau_h$,
one wins a factor of 5.5 from the branching ratio. 
Therefore the background to $\ell\tau_h\tau_h\met$
must be at least $5.5^2\sim 30$ times larger in order
for the clean trilepton channel to be still preferred.

\section{Backgrounds}\label{sec:backgrounds}

We next turn to the discussion of the backgrounds involved.
We have simulated the following physics background processes:
$ZZ$, $WZ$, $WW$, $t\bar{t}$, $Z+jets$, and $W+jets$, generating
$10^6$, $10^6$, $10^6$, $10^6$, $10^7$ and $10^7$ events, respectively.
We list the results in Tables~\ref{BKND_A}-\ref{BKND_D},
where all errors are purely statistical.
%
%
%
\begin{table*}[t!]
\centering
\renewcommand{\arraystretch}{1.5}
\begin{tabular}{||c||c|c|c|c|c||}
\hline\hline
  & \multicolumn{5}{c||}{Experimental signatures} \\ 
\cline{2-6}
  & $\ell\ell\ell\met$
    & $\ell\ell\tau_h\met$ 
      & $\ell^+\ell^+\tau_h\met$
        & $\ell\tau_h\tau_h\met$
          & $\tau_h\tau_h\tau_h\met$  \\ \hline\hline
$ZZ$                    & 0.196 $\pm$ 0.028   
                                 & 0.334 $\pm$ 0.036
                                          & 0.094 $\pm$ 0.019
                                                   & 0.181 $\pm$ 0.027
                                                            & 0.098 $\pm$ 0.020  \\ \hline
$WZ$                    & 1.058 $\pm$ 0.052   
                                 & 1.087 $\pm$ 0.053
                                          & 0.447 $\pm$ 0.034
                                                   & 1.006 $\pm$ 0.051
                                                            & 0.248 $\pm$ 0.025  \\ \hline
$WW$                    & ---   
                                 & 0.416 $\pm$ 0.061
                                          & ---
                                                   & 0.681 $\pm$ 0.078
                                                            & 0.177 $\pm$ 0.039  \\ \hline
$t\bar{t}$              & 0.300 $\pm$ 0.057   
                                 & 1.543 $\pm$ 0.128
                                          & 0.139 $\pm$ 0.038
                                                   & 1.039 $\pm$ 0.105
                                                            & 0.161 $\pm$ 0.041  \\ \hline
$Zj$                    & 0.112 $\pm$ 0.079   
                                 & 7.34 $\pm$ 0.64
                                          & 0.168 $\pm$ 0.097
                                                   & 20.3 $\pm$ 1.1
                                                            & 17.9 $\pm$ 1.0  \\ \hline
$Wj$                    & ---    & ---    & ---    & 37.2 $\pm$ 2.9
                                                            & 6.1 $\pm$ 1.2  \\ \hline\hline
$\sigma_{BG}^{\rm tot}$ 
                        & 1.67 $\pm$ 0.11   
                                 & 10.7  $\pm$ 0.7
                                          & 0.85 $\pm$ 0.11
                                                   & 60.4 $\pm$ 3.1
                                                            & 24.7 $\pm$ 1.6  \\ \hline\hline
\end{tabular}
\parbox{5.5in}{
\caption{ Results for the individual SM backgrounds (in fb), as well as the total
background $\sigma_{BG}^{\rm tot}$ in the various channels for case A:
$\met>20$ GeV and no jet veto.
\label{BKND_A}}}
\end{table*}
%
%
%
\begin{table*}[ht!]
\centering
\renewcommand{\arraystretch}{1.5}
\begin{tabular}{||c||c|c|c|c|c||}
\hline\hline  
  & \multicolumn{5}{c||}{Experimental signatures} \\ 
\cline{2-6}
  & $\ell\ell\ell\met$
    & $\ell\ell\tau_h\met$ 
      & $\ell^+\ell^+\tau_h\met$
        & $\ell\tau_h\tau_h\met$
          & $\tau_h\tau_h\tau_h\met$  \\ \hline\hline
$ZZ$                   & 0.165 $\pm$ 0.025   
                                 & 0.271 $\pm$ 0.033
                                          & 0.090 $\pm$ 0.019
                                                   & 0.153 $\pm$ 0.024
                                                            & 0.086 $\pm$ 0.018  \\ \hline
$WZ$                   & 0.964 $\pm$ 0.050   
                                 & 1.001 $\pm$ 0.051
                                          & 0.423 $\pm$ 0.033
                                                   & 0.909 $\pm$ 0.049
                                                            & 0.204 $\pm$ 0.023  \\ \hline
$WW$                   & ---   
                                 & 0.380 $\pm$ 0.058
                                          & ---
                                                   & 0.602 $\pm$ 0.073
                                                            & 0.142 $\pm$ 0.036  \\ \hline
$t\bar{t}$             & 0.300 $\pm$ 0.057   
                                 & 1.500 $\pm$ 0.127
                                          & 0.139 $\pm$ 0.038
                                                   & 0.996 $\pm$ 0.103
                                                            & 0.128 $\pm$ 0.037  \\ \hline
$Zj$                   & 0.056 $\pm$ 0.056   
                                 & 4.87 $\pm$ 0.52
                                          & 0.112 $\pm$ 0.079
                                                   & 13.61 $\pm$ 0.87
                                                            & 11.82 $\pm$ 0.81  \\ \hline
$Wj$                   & ---    & ---    & ---    & 32.1 $\pm$ 2.7
                                                            & 5.5 $\pm$ 1.1  \\ \hline\hline
$\sigma_{BG}^{\rm tot}$ 
                       & 1.49  $\pm$ 0.10
                                 & 8.0  $\pm$ 0.5
                                          & 0.76 $\pm$ 0.09
                                                   & 48.4 $\pm$ 2.8
                                                            & 17.9 $\pm$ 1.4  \\ \hline\hline
\end{tabular}
\parbox{5.5in}{
\caption{ The same as Table~\ref{BKND_A}, but for case B.
\label{BKND_B}}}
\end{table*}
%
%
%
\begin{table*}[ht!]
\centering
\renewcommand{\arraystretch}{1.5}
\begin{tabular}{||c||c|c|c|c|c||}
\hline \hline
  & \multicolumn{5}{c||}{Experimental signatures} \\ 
\cline{2-6}
  & $\ell\ell\ell\!\!\met$
    & $\ell\ell\tau_h\!\!\met$ 
      & $\ell^+\ell^+\tau_h\!\!\met$
        & $\ell\tau_h\tau_h\!\!\met$
          & $\tau_h\tau_h\tau_h\!\!\met$  \\ \hline\hline
$ZZ$                & 0.114 $\pm$ 0.021   
                                 & 0.220 $\pm$ 0.029
                                          & 0.071 $\pm$ 0.017
                                                   & 0.094 $\pm$ 0.019
                                                            & 0.031 $\pm$ 0.011  \\ \hline
$WZ$                & 0.805 $\pm$ 0.046   
                                 & 0.828 $\pm$ 0.046
                                          & 0.347 $\pm$ 0.030
                                                   & 0.695 $\pm$ 0.043
                                                            & 0.136 $\pm$ 0.019  \\ \hline
$WW$                & ---   
                                 & 0.301 $\pm$ 0.052
                                          & ---
                                                   & 0.354 $\pm$ 0.056
                                                            & 0.097 $\pm$ 0.029  \\ \hline
$t\bar{t}$          & ---   
                                 & 0.086 $\pm$ 0.030
                                          & ---
                                                   & 0.032 $\pm$ 0.018
                                                            & ---   \\ \hline
$Zj$                & 0.056 $\pm$ 0.056   
                                 & 4.93 $\pm$ 0.52
                                          & 0.056 $\pm$ 0.056
                                                   & 12.66 $\pm$ 0.84
                                                            & 10.36 $\pm$ 0.76  \\ \hline
$Wj$                & ---    & ---    & ---    & 25.8 $\pm$ 2.4
                                                            & 3.2 $\pm$ 0.9  \\ \hline\hline
$\sigma_{BG}^{\rm tot}$ 
                    & 0.97  $\pm$ 0.07
                                 & 6.4  $\pm$ 0.5
                                          & 0.47 $\pm$ 0.06
                                                   & 39.6 $\pm$ 2.5
                                                            & 13.8  $\pm$ 1.2  \\ \hline\hline
\end{tabular}
\parbox{5.5in}{
\caption{ The same as Table~\ref{BKND_A}, but for case C.
\label{BKND_C}}}
\end{table*}
%
%
%
%
\begin{table*}[ht!]
\centering
\renewcommand{\arraystretch}{1.5}
\begin{tabular}{||c||c|c|c|c|c||}
\hline\hline
  & \multicolumn{5}{c||}{Experimental signatures} \\ 
\cline{2-6}
  & $\ell\ell\ell\met$
    & $\ell\ell\tau_h\!\!\met$ 
      & $\ell^+\ell^+\tau_h\!\!\met$
        & $\ell\tau_h\tau_h\!\!\met$
          & $\tau_h\tau_h\tau_h\!\!\met$  \\ \hline\hline
$ZZ$                    & 0.098 $\pm$ 0.020   
                                 & 0.177 $\pm$ 0.026
                                          & 0.071 $\pm$ 0.017
                                                   & 0.075 $\pm$ 0.017
                                                            & 0.027 $\pm$ 0.010  \\ \hline
$WZ$                    & 0.732 $\pm$ 0.044   
                                 & 0.766 $\pm$ 0.045
                                          & 0.329 $\pm$ 0.029
                                                   & 0.622 $\pm$ 0.040
                                                            & 0.115 $\pm$ 0.017  \\ \hline
$WW$                    & ---   
                                 & 0.274 $\pm$ 0.049
                                          & ---
                                                   & 0.327 $\pm$ 0.054
                                                            & 0.071 $\pm$ 0.025  \\ \hline
$t\bar{t}$              & ---   
                                 & 0.075 $\pm$ 0.028
                                          & ---
                                                   & 0.032 $\pm$ 0.018
                                                            & ---   \\ \hline
$Zj$                    & ---   
                                 & 3.25 $\pm$ 0.24
                                          & ---
                                                   & 7.62 $\pm$ 0.65
                                                            & 6.55 $\pm$ 0.61  \\ \hline
$Wj$                    & ---    & ---    & ---    & 22.6 $\pm$ 2.3
                                                            & 3.0 $\pm$ 0.8  \\ \hline\hline
$\sigma_{BG}^{\rm tot}$ 
                        & 0.83  $\pm$ 0.05
                                 & 4.5  $\pm$ 0.3
                                          & 0.40 $\pm$ 0.03
                                                   & 31.3 $\pm$ 2.4
                                                            & 9.8  $\pm$ 1.0  \\ \hline\hline
\end{tabular}
\parbox{5.5in}{
\caption{The same as Table~\ref{BKND_A}, but for case D.
\label{BKND_D}}}
\end{table*}
A few comments are in order.
\begin{enumerate}
\item $WZ$ is indeed the major source of background for the trilepton
channel. The majority of the background events
contain a leptonically decaying off-shell $Z$ and 
pass the invariant dilepton mass cut. The rest of the WZ background comes
from $Z\rightarrow\tau^+\tau^-\rightarrow \ell^+\ell^-\met$.
The $WZ$ rate then is a factor of three higher than in
recent trilepton analyses prior to the SUSY/Higgs workshop
(see, e.g. \cite{BCDPTinPRL,BCDPTinPRD,Barger}).
To simulate the diboson backgrounds, most previous estimates 
employed ISAJET, where the $W$ and $Z$ gauge bosons are always
generated exactly on their mass shell, and there is no finite-width
smearing effect \cite{MPcomp}, \cite{MPinPRD} \footnote{Since then,
the trilepton analysis has been redone independently by several
groups and the increase in the $WZ$ background has been confirmed
\cite{MPcomp,MPinPRD,MPtalks,MPinPLB,CE,BDPQT,BK4}. In addition, the
virtual photon contribution and the $Z-\gamma$ interference effect,
neither of which is modelled in either PYTHIA or ISAJET, have also
been included \cite{MPtalks,MPinPLB,CE,BDPQT,BK4}, which further increases
the background several times. This required new cuts, specifically
designed to remove these additional contributions \cite{MPinPLB,BDPQT}.}.
\item As we move to channels with more tau jets, the number
of background events with {\em real} tau jets decreases: first,
because of the smaller branching ratios of $W$ and $Z$ to taus;
and second, because the tau jets in $W$ and $Z$ decays are {\em softer}
than the leptons from $W$ and $Z$. This is to be contrasted with
the signal, where, conversely, the tau jets are harder than the leptons.
We also see, however, that the contribution from events with
fake taus (from hadronically decaying $W$'s and $Z$'s or from initial
and final state jet radiation) increases, and for the $3\tau$
channel events with fake taus are the dominant part of the $WZ$
background.
\item Notice that the $WZ$ background to the same-sign dilepton channel 
is smaller (by a factor of two) than for the trilepton channel. As expected,
it is also about a half of the total contribution to $\ell\ell\tau$
(recall that for the signal this ratio is only a third).
Indeed, one third of the events with opposite sign leptons
come from the $Z$-decay and are cut away by the dilepton mass cut.
\item Vetoing a fourth lepton in the event reduces the $ZZ$ background
to the trilepton channel only by 4--8  \%. The $ZZ$
trilepton background is due to one $Z$ decaying as $Z\rightarrow \tau\tau$,
thus providing the missing energy in the event, and the other $Z$
decaying to leptons: $Z\rightarrow \ell^+\ell^-$. Most of the events passing
the cuts contain an off-shell $Z/\gamma$ decaying leptonically\footnote{ISAJET
analyses are missing this component of the $ZZ$ background.},
and the third lepton coming from a leptonic tau.
But then it is 6 times more probable that the
second tau would decay hadronically and will not give a fourth lepton.
The rest of the $ZZ$ background events come from a regular
$Z\rightarrow \ell^+\ell^-$ decay, where one of the leptons is missed,
and the invariant mass cut does not apply. For those events, there
is obviously no fourth lepton.
\item The jet veto is very effective in reducing the $t\bar{t}$ background
for the first three channels. However, it also reduces the signal
(see Fig.~\ref{sigeff}).
\item In all channels, a higher $\met$ cut did not
help to get rid of the major backgrounds. Indeed, $WZ$,
$t\bar{t}$ and/or $Wj$ backgrounds tend to have a lot of missing energy,
due to the leptonic $W$-decays.
\item Our result for the $Wj$ and $Zj$ backgrounds should be taken
with a grain of salt, in spite of the relatively small statistical errors.
Events with fake leptons are expected to comprise a major part of this
background, and SHW does not provide a realistic simulation of those. 
In fact, the most reliable way to estimate this background will
be from Run IIa data, e.g. by estimating the probability for an
isolated track from Drell-Yan events, and the lepton fake rate
per isolated track from minimum bias data \cite{JN,MPinPRD,MPinPLB}.
\item We have underestimated the total background
to the three-jet channel by considering only processes with at least
one real tau in the event. We expect sizable contributions from
pure QCD multijet events, or $Wj\rightarrow jjj$, where
{\em all} three tau jets are fake. 
\end{enumerate}

\section{Triggers} \label{sec:triggering}

Since the four experimental signatures in our analysis contain only
soft leptons and tau jets, an important issue is whether one can
develop efficient combinations of Level 1 and Level 2 triggers to
accumulate these data sets without squandering all of the available bandwidth.
A dedicated low $p_T$ tau trigger for Run II, which may be suitable
for the new tau jet channels, is now being considered by CDF
\cite{tau trigger}.

In order to get an idea how well we can trigger on these new channels
in Run II, we made use of the existing trigger objects in the SHW package.
Until the design and approval of a dedicated tau trigger, which will
collect most of the signal sample by itself, we want to make sure 
that the signal events will somehow end up on tape with the already
existing triggers and will not be lost. This is why we considered
a standard set of triggers used for SUSY analyses at the Tevatron
\cite{BCDPTinPRL,BCDPTinPRD}. Of the five triggers used in \cite{BCDPTinPRL,BCDPTinPRD}
we discarded the multijet trigger as not useful for our channels,
and the dilepton plus $\met$ trigger as unrealistic for Run II.
We then conservatively tightened the thresholds of the
$\met$ trigger and the single lepton trigger:
\begin{enumerate}
\item $\met>40$ GeV;
\item $p_T(\ell)>20$ GeV;
\item $p_T(\ell)>10$ GeV, $p_T(j)>15$ GeV and $\met>15$ GeV,
\end{enumerate}
and counted how many of the signal events {\em after cuts}
were collected by these three triggers. The results are shown in
Table~\ref{triggers}.
\begin{table*}[t!]
\centering
\renewcommand{\arraystretch}{1.5}
\begin{tabular}{||c||c|c|c|c||}\hline\hline
{\small $m_{\tilde \chi^\pm_1}$ (GeV)}
  & $\ell\ell\ell\met$ 
    & $\ell\ell\tau\met$
      & $\ell\tau\tau\met$
        & $\tau\tau\tau\met$  \\
\hline\hline
100 & 82\% & 94\% & 83\% & 50\%  \\ \hline
110 & 85\% & 93\% & 83\% & 52\%  \\ \hline
123 & 95\% & 91\% & 86\% & 55\%  \\ \hline
140 & 93\% & 95\% & 89\% & 60\%  \\ \hline\hline
\end{tabular}
\parbox{5.5in}{
\caption{Fraction of signal events {\em after cuts}
collected by the set of three triggers described in the text.
\label{triggers}}}
\end{table*}
We can see that across the board the three very simplified triggers
did a very good job and typically picked up about 90\% of the signal events
which contained at least one lepton. 

We also checked if we can use the tau trigger in SHW 
(which is calorimetry-based and probably not the most
suitable trigger for our purposes \cite{tau trigger})
to collect some of the remaining events. We considered the effect
of a mixed $\ell+\tau$ trigger and found only some marginal improvement
of a few percent. The only case, in which a new trigger helps a lot is the
jetty channel $\tau_h\tau_h\tau_h$, where it is a $\tau-\tau$ and not
$\ell-\tau$ trigger which is relevant. However, developing a stand-alone
hadronic double tau trigger does not seem well justified -- the fake rate
for QCD jets faking taus is large enough to make the trigger fire mostly
on pure QCD events, where two jets are faking taus. In order to even entertain
the idea of a double tau trigger, one would have to think seriously about
adding an extra requirement, for example $\met$, but then the trigger
becomes too complicated. Besides, it helps a lot only for the channel with
the worst reach (and largest backgrounds). 
It may in fact be a better idea to lower the threshold of the
$\met$ trigger instead. 

\section{Tevatron Reach} \label{sec:reach}

A $3\sigma$ exclusion limit would require a total integrated luminosity
\begin{equation}
L\ =\ {9\sigma_{BG}\over 
\left(\sigma_{sig}\ 
BR(\tilde\chi^+_1\tilde\chi^0_2\rightarrow X)\ \epsilon_{tot}\right)^2}.
\end{equation}
Notice that $L(3\sigma)$ depends linearly on the
background $\sigma_{BG}$ after cuts, but {\em quadratically} on
the signal branching ratios. This allows the jetty channels
to compete very successfully with the clean trilepton signature,
whose branching ratio is quite small (see Table~\ref{tau_branching}).
\begin{figure}[t!]
\epsfysize=3.0in
\epsffile[-30 215 290 545]{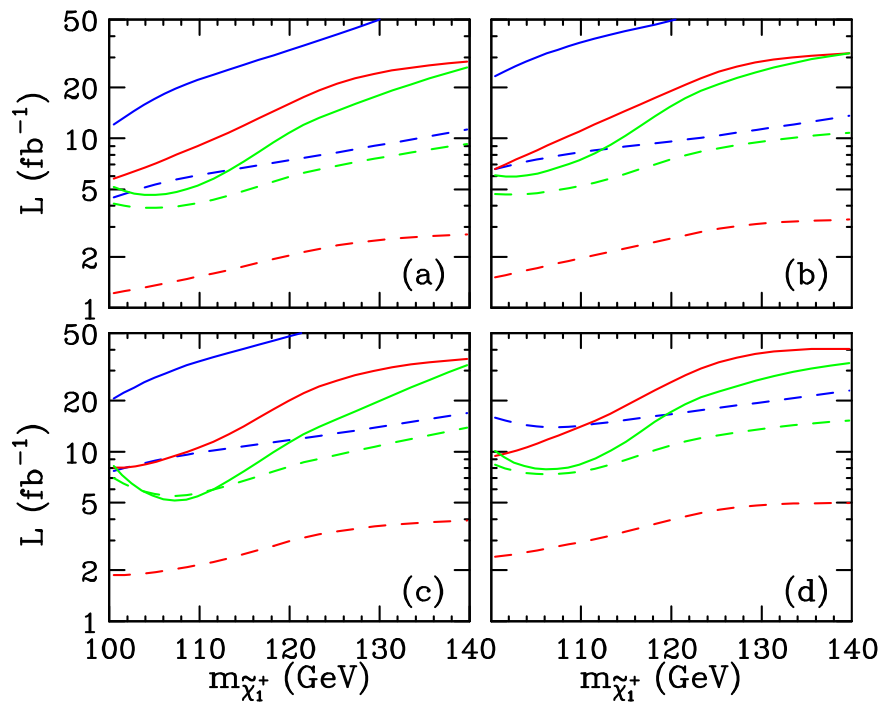}
\begin{center}
\parbox{5.5in}{
\caption[]{ The total integrated luminosity $L$ needed
for a $3\sigma$ exclusion (solid lines) or observation of
5 signal events (dashed lines), as a function of the chargino
mass $m_{\tilde\chi^+_1}$, for the three channels:
$\ell\ell\ell\met$ (blue), $\ell\ell\tau_h\met$ (red) and
$\ell^+\ell^+\tau_h\met$ (green); and 
for various sets of cuts: (a) cuts A; (b) cuts B;
(c) cuts C and (d) cuts D.
\label{reach}}}
\end{center}
\end{figure}
In Fig.~\ref{reach} we show the Tevatron reach in the three
channels: trileptons (blue), dileptons plus a tau jet (red)
and like-sign dileptons plus a tau jet (green).
We see that the two channels with tau jets have a much better
sensitivity compared to the usual trilepton signature.
Assuming that efficient triggers can be implemented,
the Tevatron reach will start exceeding LEP II
limits as soon as Run IIa is completed and the two collaborations
have collected a total of $4\ {\rm fb}^{-1}$ of data.
Considering the intrinsic difficulty of the SUSY scenario
we are contemplating, the mass reach for Run IIb is quite
impressive. One should also keep in mind that
we did not attempt to optimize our cuts for the new channels.
For example, one could use angular correlation cuts to suppress Drell-Yan,
transverse $W$ mass cut to suppress $WZ$ \cite{BDPQT},
or (chargino) mass--dependent $p_T$ cuts for the leptons and tau jets
\cite{MPinPRD,MPinPLB}, to squeeze out some extra reach.
In addition, the $\ell\ell\tau_h$ channel
can be explored at smaller values of $\tan\beta$ as well
\cite{BCDPTinPRD,Barger,MPinPRD,MPinPLB}, since the two-body
chargino decays are preferentially to tau sleptons.
In that case, the clean trilepton channel still offers the best reach,
and a signal can be observed already in Run IIa. Then,
the tau channels will not only provide an important confirmation,
but also hint towards some probable values of the SUSY model parameters.

\vspace{.25in}

{\bf Acknowledgements.}
We would like to thank V.~Barger, J.~Conway, R.~Demina, L.~Groer,
J.~Nachtman, D.~Pierce, A.~Savoy-Navarro and M.~Schmitt for
useful discussions.
Fermilab is operated under DOE contract DE-AC02-76CH03000.

\end{document}